\documentclass[final,5p,times,twocolumn]{elsarticle}

\usepackage{amssymb}
\usepackage{array}
\usepackage{pifont}
\usepackage[colorlinks=true, breaklinks=true, linkcolor=red, citecolor=blue, urlcolor=blue]{hyperref} 
\usepackage{hyperref}

\begin{document}

\begin{frontmatter}



\title{GPTArticleExtractor: An Automated Workflow for Magnetic Material Database Construction\tnoteref{1}}


\author[a,b]{Yibo Zhang}
\ead{yibo.zhang@unh.edu}
\author[a]{Suman Itani}
\author[a]{Kamal Khanal}
\author[a]{Emmanuel Okyere}
\author[a]{Gavin Smith}
\author[a]{Koichiro Takahashi}
\author[a]{Jiadong Zang}
\ead{jiadong.zang@unh.edu}

\affiliation[a]{organization={Department of Physics and Astronomy, University of New Hampshire},
            addressline={9 Library Way}, 
            city={Durham},
            postcode={03824}, 
            state={NH},
            country={USA}}

\affiliation[b]{organization={Department of Chemistry, University of New Hampshire},
            addressline={23 Academic Way}, 
            city={Durham},
            postcode={03824}, 
            state={NH},
            country={USA}}

\begin{abstract}
A comprehensive database of magnetic materials is valuable for researching the properties of magnetic materials and discovering new ones.
This article introduces a novel workflow that leverages large language models for extracting key information from scientific literature. From 22,120 articles in the Journal of Magnetism and Magnetic Materials, a database containing 2,035 magnetic materials was automatically generated, with ferromagnetic materials constituting 76\% of the total.
Each entry in the database includes the material's chemical compounds, as well as related structures (space group, crystal structure) and magnetic temperatures (Curie,  N\'eel, and other transitional temperatures). To ensure data accuracy, we meticulously compared each entry in the database against the original literature, verifying the precision and reliability of each entry.


\end{abstract}




\end{frontmatter}


\section{Introduction}
\label{Introduction}

Magnetic materials play an indispensable role in modern science and engineering domains. Their applications span a wide range, from data storage devices like hard disks and tapes to electrical power conversion and transmission systems, and extend to medical and consumer electronics. The unique magnetic properties of these materials enable various technological applications to function efficiently and precisely \cite{coey2012magnetism, spaldin2010magnetic}.

While discovering antiferromagnets is relatively easy due to the superexchange nature therein\cite{anderson1950antiferromagnetism}, predicting new ferromagnets poses significant challenges. The interplay of the correlation and itinerancy is believed to be the origin of most ferromagnetism, but a complete understanding is yet to be available \cite{Tasaki2020}. The first-principles calculations, such as density functional theory (DFT), are generally accurate in determining the energy of non-magnetic materials. Unfortunately, they often result in incorrect predictions of magnetic properties, especially for magnetic materials where the electronic correlation is relevant, particularly ferromagnetism. Many magnetic materials are strongly correlated, and one often has to go beyond DFT or DFT + U approaches \cite{Pavarini2021}.
Moreover, the limitations of the DFT+U method in predicting magnetic structures are demonstrated by the disagreement between the observed experimental and the calculated magnetic configurations \cite{Blanco2009}. Another drawback of the DFT calculations is that to get a successful description of exchange interactions in some materials (like antiferromagnetic Mn$_3$O$_4$), one has to use hybrid exchange-correlation functionals, which have a considerable computational cost \cite{Ribeiro2015}. Another study \cite{Romero2018} also suggests that none of the functionals work in all conditions, indicating the need for further development of exchange-correlation functionals. The paramount impact is clearly doomed if a reliable way of predicting new ferromagnets cannot be achieved.

Data-driven materials discovery is an effective solution to this challenge. For example, the high-throughput (HT) methods efficiently pinpoint materials with distinct properties by conducting broad screenings within databases \cite{curtarolo2013high,materialsproject, liu2023atomly}. However, HT outcomes, rooted in the first-principles calculations, are still constrained by these algorithms' drawbacks above.
On the other hand, Data mining techniques in materials science are significantly enhanced by various analytical methods. Key among these are machine learning \cite{nelson2019predicting, singh2023physics}, deep learning \cite{alverson2023generative}, linear regression \cite{nguyen2019regression}, and trend analysis \cite{byland2022statistics}, which all garnered success in identifying materials and developing screening models. However, the effectiveness of these methods relies heavily on the availability of a comprehensive dataset, which has yet to be ready for magnetic materials.

Databases based on experimentally verified information do exist \cite{vaitkus2021validation,zagorac2019recent,ICSD,pearson}, yet they all lack magnetic property information, such as Curie or N\'eel temperatures. The Bilbao Magnetic Materials Database \cite{gallego2016magndata1} comes closest to meeting these criteria. It offers a complete lattice and magnetic structure for each entry and leads to success in discovering magnetic topological insulators \cite{xu2020high}. However, its dataset is limited to 1,890 entries, a scale insufficient to support more intricate deep-learning models. Some efforts to create a magnetic material database have been noticed in articles \cite{nelson2019predicting, singh2023physics, xu2011inorganic} and books\cite{connolly2012bibliography,buschow2003handbook,villars2004data}. Notably, Matthew et al. \cite{swain2016chemdataextractor} developed a rule-based syntax toolkit for the automatic extraction of chemical information. It requires substantial programming efforts to use regular expressions, which may fail to adapt to different articles' writing styles and layouts. Callum et al. \cite{court2018auto} automated the analysis of 68,078 articles in physics and chemistry, extracting information related to chemical composition and magnetic phase transitions (e.g., Curie and N\'eel temperatures) from 39,822 of them. Luke et al. \cite{gilligan2023rule} implemented literature scraping without rules by fine-tuning a BERT model. These works do not address the extraction of structural information, which is essential for the descriptor \cite{himanen2020dscribe} in data-based materials discovery. However, we admit the difficulty in extracting structural information since they are often embedded within extensive text. Furthermore, these works do not distinguish experimental and theoretical papers, risking contamination of experimental data with theoretical data in the database. Data quality needs to be further evaluated.

The recent development of large language models (LLM) in natural language processing (NLP) brings new insight into this challenge. It can be traced back to the advent of the Transformer model \cite{vaswani2017attention}. This model introduced the Self-Attention mechanism, laying the groundwork for NLP. Further advances came with the pre-trained BERT model \cite{devlin2018bert}, solidifying the field's foundation. OpenAI subsequently followed suit by releasing their version of a pre-trained model \cite{brown2020language}. This technology's widespread application and significant advancement are attributed to OpenAI's theory of fine-tuning\cite{ouyang2022training}. This theory empowers LLMs to understand human intent and engage in meaningful dialogue and interaction based on these intents. It makes machine reading of scientific literature possible.



The process of training LLMs generally consists of two main steps. The first step, the pre-training \cite{devlin2018bert,brown2020language,touvron2023llama2}, involves embedding human knowledge and linguistic structures into the model's internal parameters, similar to lossy compression \cite{deletang2023language} of this knowledge in the model's memory. The second step, Fine-Tuning, particularly through the Instruction Tuning\cite{touvron2023llama2}, transforms a pre-trained large language model into a chat model capable of effective human interaction.

However, the model's effectiveness is limited by its training methods. It performs well in areas with ample data but struggles in less-explored fields. Additionally, as it is not trained specifically for literature extraction, using it to convert magnetic material texts directly into a database could be ineffective. These limitations highlight the importance of precise instructions and prompt engineering \cite{zhou2022large}.

Prompt Engineering enhances interactions with LLMs by designing effective prompts to guide more accurate outputs, even when the model has limited knowledge in areas such as magnetic materials or when instruction tuning is not specifically optimized for certain tasks.

In this work, through prompt engineering of LLM, we have developed a novel method named GPTArticleExtractor for extracting bibliographic data to create a database. To validate the feasibility of our approach, we have constructed a database for magnetic materials based on published papers in the Journal of Magnetism and Magnetic Materials.

\begin{figure*}[ht]
	\begin{center}
		\includegraphics[width=\textwidth]{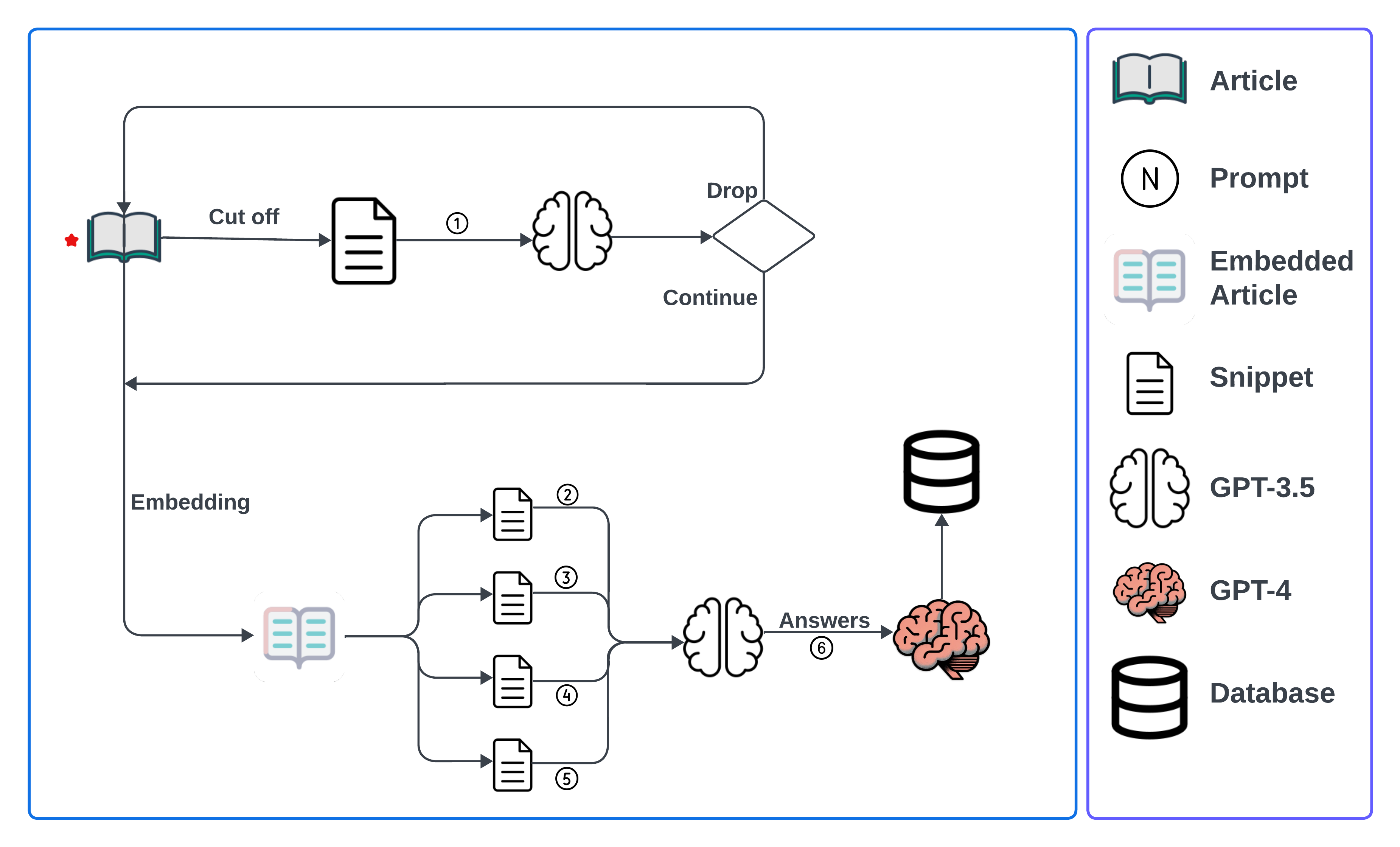}
        \vspace{-30pt}
		\caption{Flowchart depicting the transformation from original articles to structured data through various prompts and processing stages, utilizing different GPT versions.}
		\label{fig:workflow}
	\end{center}
\end{figure*}


\begin{table}[h]
	\small
    \centering
    \caption{Prompts}
    \label{tab:task_list}
    \begin{tabular}{|m{2cm}|m{6cm}|}
        \hline
        Task & Prompt \\
        \hline
        \ding{172} Study Type \newline \hspace*{0.8em} Filter  & Based on the provided abstract and title of an article, determine if this article is a research study about materials. Please return 'True' if it is and 'False' if it isn't. \\
        \hline
        \ding{173} Temperature & Please determine the Curie temperature, N\'eel temperature, transitional temperature or critical temperature discussed in this article. Please include the specific name of these temperature. Consider the following documents: \{doc\_tem\} \\
        \hline
        \ding{174} Materials & Please identify the chemical compounds being studied in this article. Consider the following documents: \{doc\_material\} \\
        \hline
        \ding{175} Structure & Please help me find and summarize the paragraphs related to material structure, crystal structure, lattice structure, and other relevant concepts in the given text: \{doc\_structure\} \\
        \hline
        \ding{176} Space Group & Please help me find and summarize the paragraphs discussing space groups, their properties, and their relevance to crystal structures in the given text: \{doc\_space\_group\} \\
        \hline
        \ding{177} Summary & As a research assistant, your task is to identify the material being studied in this article, its associated Curie temperature, transition temperatures and critical temperatures, and structural information like space group, lattice Constant, crystal structure, lattice structure etc., based on the following details provided to you: the title, abstract, and summary details from GPT. The title and abstract should be your primary sources, while GPT's summaries should serve as secondary references. Title: \{title\}, Abstract: \{abstract\}, GPT Materials Summary: \{materials\}, GPT Temperature Summary: \{temperature\}, GPT Structural Information: \{structure\}. The output should be in a JSON readable format. \\
        \hline
    \end{tabular}
\end{table}

\section{Material and methods}

This section presents a comprehensive overview of the workflow employed in GPTArticleExtractor. Our principal aim is to extract critical information pertaining to each experimentally reported material, including the material's chemical formula, various magnetic temperatures (such as Curie, Curie-Weiss, N\'eel, and other transitional temperatures), as well as structural details like the space group and crystal structure.


\subsection{Data and Tools}

\textbf{DATA}: This study utilizes Elsevier's Text Mining Application Programming Interface (API) for text retrieval through DOI queries. It focuses on articles from the Journal of Magnetism and Magnetic Materials (JMMM) from 2000 to 2023. Each article is provided in Extensible Markup Language (XML) format. A substantial database comprising 22,120 corpora was downloaded.

\textbf{Model}: In this work, we employ two major language models, GPT-3.5 and GPT-4, for the automated extraction of structured data from scientific literature. Each model has unique strengths: GPT-3.5 (\$0.002 per 1K tokens) is effective for initial filtering and categorization, while GPT-4 (\$0.06 per 1K tokens) excels at deeper analysis and summarization. To ensure both efficiency and cost-effectiveness, the majority of the tasks are handled by GPT-3.5. For the final summarization phase, we switch to GPT-4 to ensure quality.


\textbf{Vector Database}: Our method is initiated by tokenizing articles segmenting the text into chunks of 500 tokens each. We then analyze these segments in a vector space, employing metrics such as Euclidean distance to measure word similarity. This analysis involves comparing the vectorized text segments with a question to determine similarity and identify the most relevant article segments for the question. The five most closely related segments to a query are identified using Facebook AI Similarity Search (FAISS)\cite{johnson2019billion}. This approach allows us to concentrate on the most pertinent segments, significantly enhancing the quality of our answers by focusing on relevance and minimizing input size.


\subsection{GPTArticleExtractor Structure}



Our platform offers a streamlined, automated text analysis process that accepts plain text and produces structured data. Figure \ref{fig:workflow} illustrates the complete workflow.
The corresponding prompts for each step are listed in Table \ref{tab:task_list}.

First, we extract the title and abstract from an article and use GPT-3.5, guided by question \ding{172}, for an initial screening to identify articles focused on new material research. If the model deems it relevant (judged as True), we proceed to the next step: vectorizing the article using a vector database. This vectorization allows us to find the most relevant snippets based on specific sub-questions (\ding{173}, \ding{174}, \ding{175}, \ding{176}). These snippets and the questions are fed back into GPT-3.5, which generates corresponding answers. This step can also be regarded as a form of text summary. Finally, these answers, along with question \ding{177}, are fed into GPT-4 for answer aggregation and structuring, which are then inserted into the database.

\section{Results}

\begin{figure}[h]
	\begin{center}
		\includegraphics[width=\linewidth]{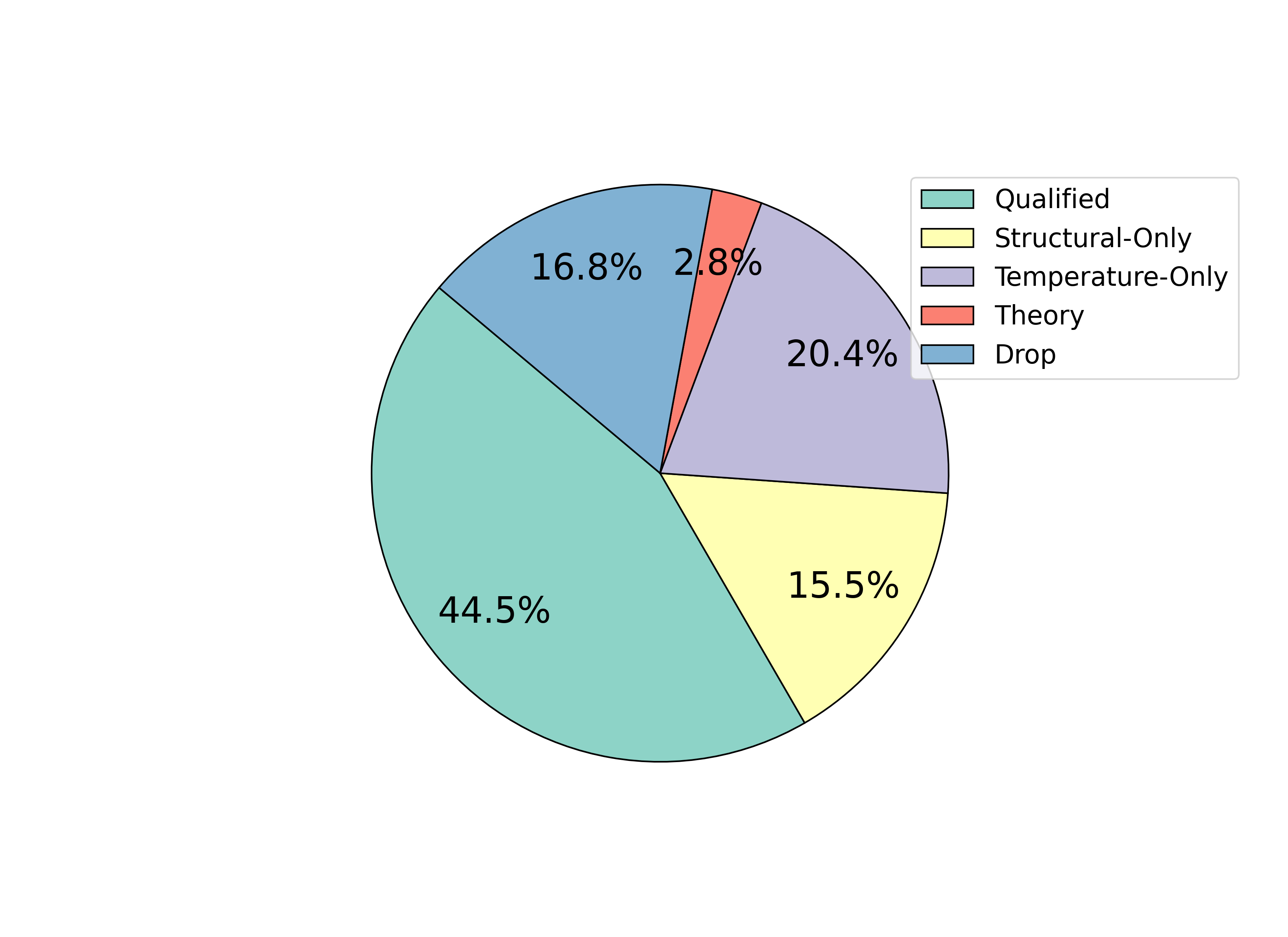}
        \vspace{-50pt}
		\caption{Data Completeness Pie Chart - 44.5\% entries ready, 15.5\% contain structure only, 20.4\% contain temperature only, 2.8\% theory, 16.8\% irrelevant. Finalized database: 2,035 entries out of 4,639 articles, yielding an accuracy rate of 83.2\%.}
		\label{fig:pie_category}
	\end{center}
\end{figure}

Using GPTExtractor, 4,639 articles were identified to report magnetic materials. To evaluate the algorithm's effectiveness, we manually checked the GPT-generated entry with the original article. As shown in Figure \ref{fig:pie_category}, 44.5\% entries have complete structural and transition temperature information. These entries are ready to be included in the database. 15.5\% and 20.4\% are attributed to structure and temperature only, respectively. These articles may be used as backup options and could be incorporated into the final database once the missing information is supplied. Noticeably, 2.8\% entries are extracted from theory papers only. Most transition temperatures therein are computed by Monte Carlo simulations or combined with the first-principles calculations. Since only experimentally verified data are included in this database, they have to be discarded. Finally, the remaining 16.8\% entries are completely irrelevant. Some of them are non-magnetic materials. Our finalized database comprises 2,035 entries. The overall 83.2\% yield rate shows the accuracy of the algorithm. This suggests that constructing a larger database focused on magnetic materials is viable once the pool of journal articles is expanded.

\begin{figure}[h]
	\begin{center}
		\includegraphics[width=\linewidth]{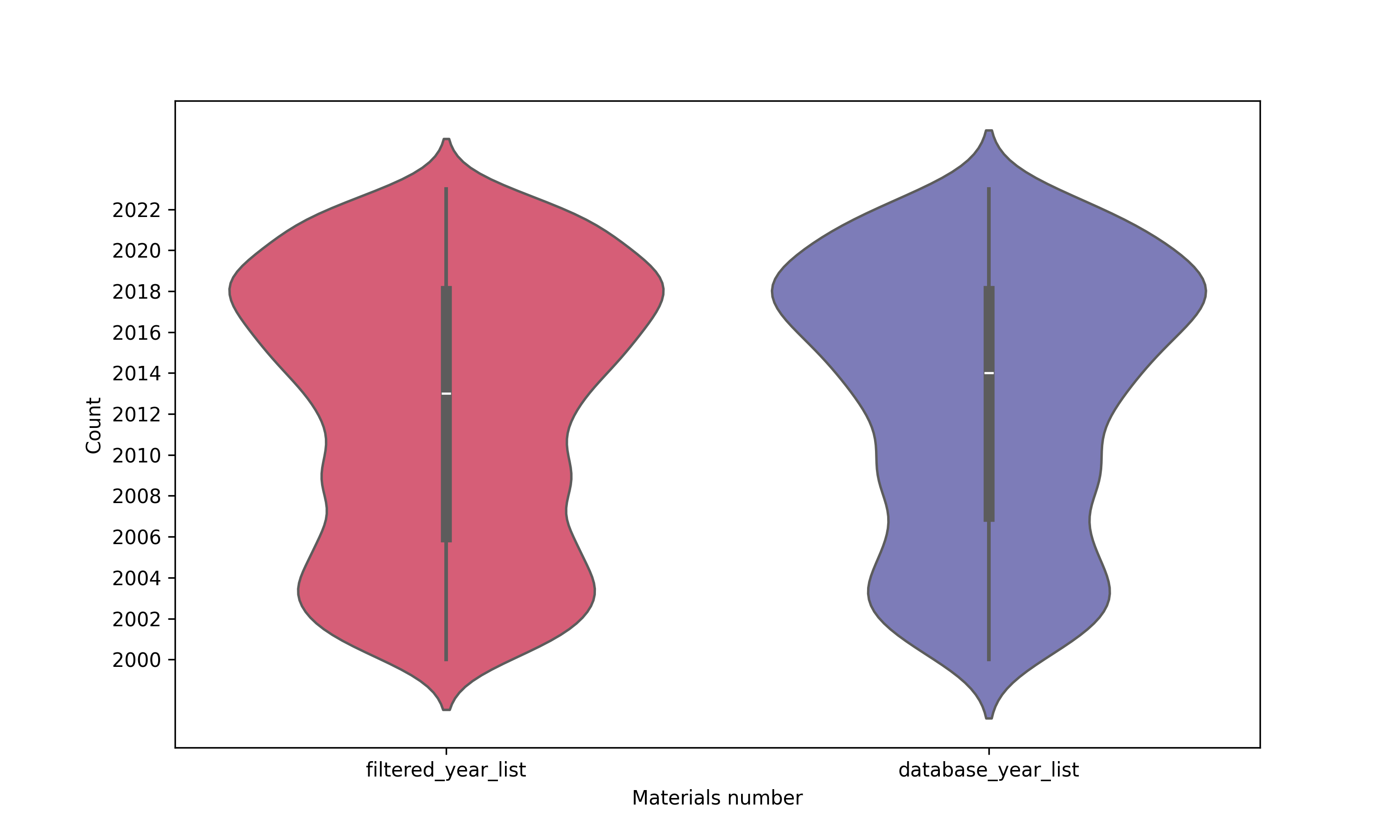}
        \vspace{-30pt}
		\caption{Violin plots comparing the Curie temperature distribution over the years for two datasets: filtered and complete database records.}
		\label{fig:violin_year}
	\end{center}
\end{figure}

Violin plots in Figure \ref{fig:violin_year} show the distribution of publication years of the articles identified. The plot on the left represents the distribution of publication years for articles in the database screened by workflow. In contrast, the plot on the right shows the distribution of articles eventually listed in the database. It is evident that a significant number of articles published before approximately 2012 were excluded during the manual review process. This observation suggests that the model is more proficient at comprehending articles from the last decade and is less susceptible to errors. Articles published before 2000 are notably absent since they are not analyzed in this work. Those articles are less commonly preserved in text form but often exist in PDF format only. Incorporating Optical Character Recognition (OCR) \cite{wick2018calamari} in future work could remedy this limitation.

\begin{figure}[h]
	\begin{center}
		\includegraphics[width=\linewidth]{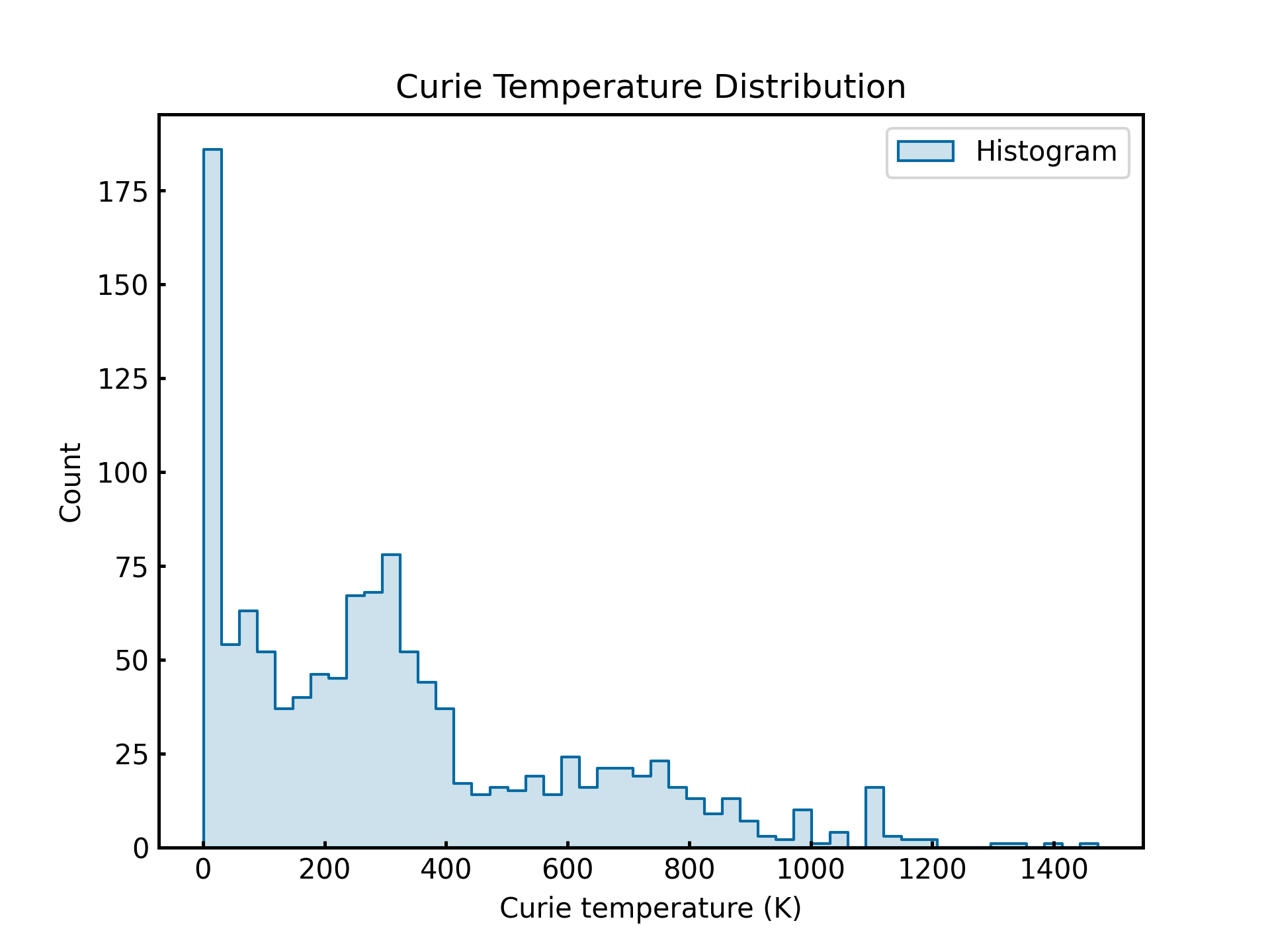}
        \vspace{-25pt}
		\caption{Histogram depicting the distribution of Curie temperatures within our comprehensive database of magnetic materials}
		\label{fig:curie_temp_distribution}
	\end{center}
\end{figure}

\begin{figure}[h]
	\begin{center}
		\includegraphics[width=\linewidth]{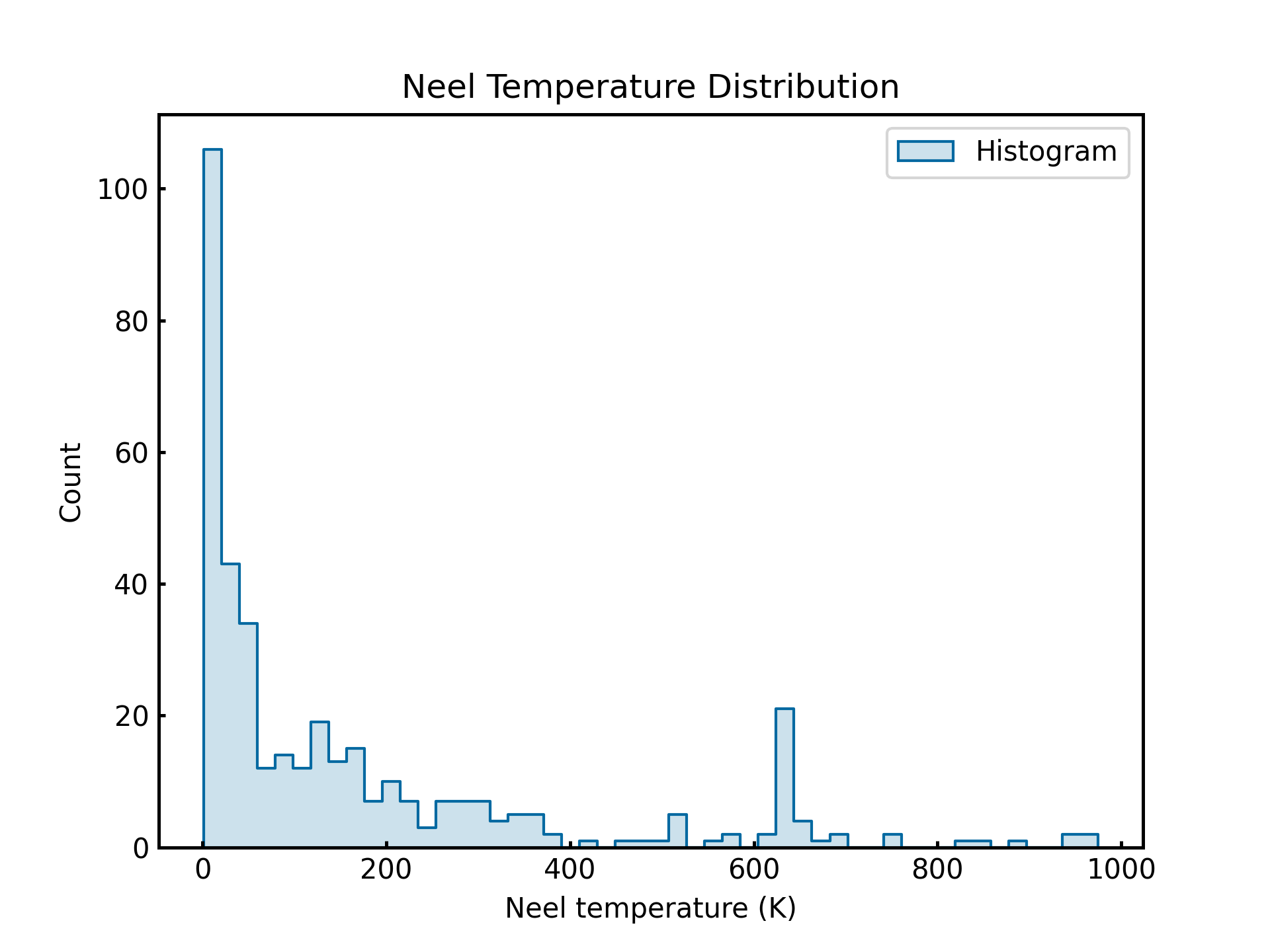}
        \vspace{-25pt}
		\caption{Histogram depicting the distribution of N\'eel temperatures within our comprehensive database of magnetic materials.}
		\label{fig:neel_temp_distribution}
	\end{center}
\end{figure}

\begin{figure}[h]
	\begin{center}
		\includegraphics[width=\linewidth]{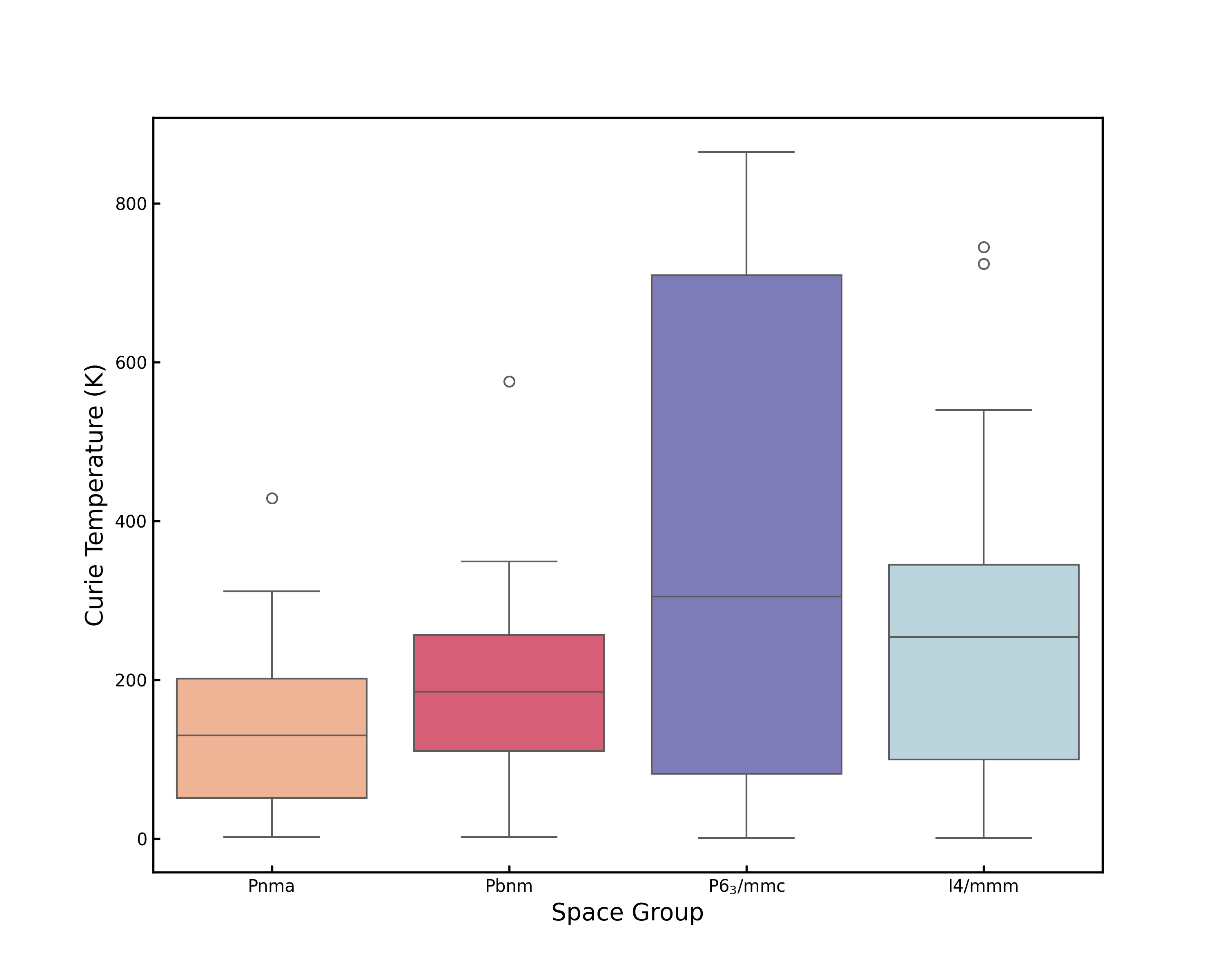}
        \vspace{-30pt}
		\caption{ Box plot displaying the distribution of Curie temperatures across different space groups in our database. Each box represents the interquartile range (IQR) and median, and outliers are indicated by individual points.}
		\label{fig:curie_vs_space_group}
	\end{center}
\end{figure}

Concerning Curie temperatures, as illustrated in Figure \ref{fig:curie_temp_distribution}, the distribution of this database spans from 0 to 1,500 K. Notably, a significant concentration of data points resides in the low-temperature range between 0 and 400 K, whereas the high-temperature region is relatively underrepresented. As for N\'eel temperatures, Figure \ref{fig:neel_temp_distribution} shows that the principal distribution ranges from 0 to 1,000 K, revealing a greater prevalence of data in the low-temperature area. Additionally, we investigated the influence of different space groups on the distribution of Curie temperatures. Figure \ref{fig:curie_vs_space_group} reveals the significant impact of diverse space groups on Curie temperatures. The \textit{P63/mmc} space group notably exhibits a broader distribution, especially at higher temperatures. Across various structures, notable contributions from Perovskite include 18\% in \textit{Pbnm}, 10\% in \textit{Pnma}, and 5.2\% in \textit{I4/mmm}. Wurtzite constitutes 5.8\% in \textit{P63/mmc}, and Spinel structures account for 1.3\% in \textit{I4/mmm}. Calculating ratios for the entire database further highlights these correlations: Perovskite at 10.9\%, Spinel at 8.5\%, and Wurtzite at 0.64\%. This data underscores a discernible link between crystal structure and space group, providing valuable insights into understanding Curie temperatures.



After constructing the database, we randomly selected 100 articles that the GPT model for manual check had not initially identified. We discovered that 13\% of these articles fulfilled the criteria for inclusion in the database. Relevant sentences in these articles are usually long and complex, posing challenges to the language model's ability to extract answers. The 500-token limit during the embedding process could truncate such sentences, leading to incomplete source data for the GPT model. Additionally, information regarding the temperature of magnetic materials is often presented in images or tables, which the GPT model could not interpret due to formatting constraints. Based on the error rate, about 2,000 additional articles are potentially eligible to be included in the database.

Several attempts could be explored in future work to address these limitations and include all eligible entries to the database. First, given the recent upgrades to the GPT model with the increasing token limit, the length of truncated embeddings could be extended to ensure answers are not prematurely cut off. The prompts could be refined, and additional few-shot learning samples could be used to accommodate articles that use diverse language styles. Future efforts could employ OCR to transcribe text from visual content to address issues associated with images, PDFs, and tables. Multimodal language models may also be leveraged to recognize and convert image-based content into text, which can then be processed further by the GPT model. 


\section{Conclusions}

GPTArticleExtractor provides an efficient way to automatically extract information about chemical compounds, magnetic temperature (Curie and N\'eel temperatures), and material structures. This tool is highly scalable and can be easily adapted for multiple database domains by simply changing the input queries. Its utility in magnetic materials is particularly notable, offering a user-friendly alternative to more complex extraction methods. A database covering various magnetic materials is available on our website (\href{https://MagneticMaterials.org}{https://MagneticMaterials.org}). The database will be constantly updated and expanded.


\section*{Acknowledgments}
This work was supported in part by a Collaborative Research Excellence (CoRE) grant from the University of New Hampshire.



\bibliographystyle{elsarticle-num} 
\bibliography{reference}

\begin{thebibliography}{10}
\expandafter\ifx\csname url\endcsname\relax
  \def\url#1{\texttt{#1}}\fi
\expandafter\ifx\csname urlprefix\endcsname\relax\def\urlprefix{URL }\fi
\expandafter\ifx\csname href\endcsname\relax
  \def\href#1#2{#2} \def\path#1{#1}\fi

\bibitem{coey2012magnetism}
J.~M.~D. Coey, Magnetism and magnetic materials, Cambridge university press, 2012.

\bibitem{spaldin2010magnetic}
N.~A. Spaldin, Magnetic materials: fundamentals and applications, Cambridge university press, 2010.

\bibitem{anderson1950antiferromagnetism}
P.~W. Anderson, Antiferromagnetism. theory of superexchange interaction, Physical Review 79~(2) (1950) 350.

\bibitem{Tasaki2020}
H.~Tasaki, The Origin of Ferromagnetism, Springer International Publishing, Cham, 2020, pp. 371--455.

\bibitem{Pavarini2021}
E.~Pavarini, Solving the strong-correlation problem in materials, La Rivista del Nuovo Cimento 44~(11) (2021) 597--640.

\bibitem{Blanco2009}
J.~A. Blanco, P.~J. Brown, Comment on dft+u search for the energy minimum among eight collinear and noncollinear magnetic structures of gdb$_4$, Phys. Rev. B 79 (2009) 216401.

\bibitem{Ribeiro2015}
R.~Ribeiro, S.~{de Lazaro}, S.~Pianaro, Density functional theory applied to magnetic materials: Mn3o4 at different hybrid functionals, Journal of Magnetism and Magnetic Materials 391 (2015) 166--171.

\bibitem{Romero2018}
A.~H. Romero, M.~J. Verstraete, From one to three, exploring the rungs of jacob's ladder in magnetic alloys, The European Physical Journal B 91~(8) (2018) 193.

\bibitem{curtarolo2013high}
S.~Curtarolo, G.~L. Hart, M.~B. Nardelli, N.~Mingo, S.~Sanvito, O.~Levy, The high-throughput highway to computational materials design, Nature materials 12~(3) (2013) 191--201.

\bibitem{materialsproject}
A.~Jain, S.~P. Ong, G.~Hautier, W.~Chen, W.~D. Richards, S.~Dacek, S.~Cholia, D.~Gunter, D.~Skinner, G.~Ceder, et~al., Commentary: The materials project: A materials genome approach to accelerating materials innovation, APL materials 1~(1) (2013).

\bibitem{liu2023atomly}
M.~Liu, S.~Meng, Atomly. net materials database and its application in inorganic chemistry, Sci Sin-Chim 53 (2023) 19--25.

\bibitem{nelson2019predicting}
J.~Nelson, S.~Sanvito, Predicting the curie temperature of ferromagnets using machine learning, Physical Review Materials 3~(10) (2019) 104405.

\bibitem{singh2023physics}
P.~Singh, T.~Del~Rose, A.~Palasyuk, Y.~Mudryk, Physics-informed machine-learning prediction of curie temperatures and its promise for guiding the discovery of functional magnetic materials, Chemistry of Materials 35~(16) (2023) 6304--6312.

\bibitem{alverson2023generative}
M.~Alverson, S.~Baird, R.~Murdock, J.~Johnson, T.~Sparks, et~al., Generative adversarial networks and diffusion models in material discovery (2023).

\bibitem{nguyen2019regression}
D.-N. Nguyen, T.-L. Pham, V.-C. Nguyen, A.-T. Nguyen, H.~Kino, T.~Miyake, H.-C. Dam, A regression-based model evaluation of the curie temperature of transition-metal rare-earth compounds, in: Journal of Physics: Conference Series, Vol. 1290, IOP Publishing, 2019, p. 012009.

\bibitem{byland2022statistics}
J.~K. Byland, Y.~Shi, D.~S. Parker, J.~Zhao, S.~Ding, R.~Mata, H.~E. Magliari, A.~Palasyuk, S.~L. Bud'ko, P.~C. Canfield, et~al., Statistics on magnetic properties of co compounds: A database-driven method for discovering co-based ferromagnets, Physical Review Materials 6~(6) (2022) 063803.

\bibitem{vaitkus2021validation}
A.~Vaitkus, A.~Merkys, S.~Gra{\v{z}}ulis, Validation of the crystallography open database using the crystallographic information framework, Journal of applied crystallography 54~(2) (2021) 661--672.

\bibitem{zagorac2019recent}
D.~Zagorac, H.~M{\"u}ller, S.~Ruehl, J.~Zagorac, S.~Rehme, Recent developments in the inorganic crystal structure database: theoretical crystal structure data and related features, Journal of applied crystallography 52~(5) (2019) 918--925.

\bibitem{ICSD}
G.~Bergerhoff, I.~Brown, F.~Allen, et~al., Crystallographic databases, International Union of Crystallography, Chester 360 (1987) 77--95.

\bibitem{pearson}
P.~Villars, K.~Cenzual, et~al., Pearson's crystal data: crystal structure database for inorganic compounds, (No Title) (2007).

\bibitem{gallego2016magndata1}
S.~V. Gallego, J.~M. Perez-Mato, L.~Elcoro, E.~S. Tasci, R.~M. Hanson, K.~Momma, M.~I. Aroyo, G.~Madariaga, Magndata: towards a database of magnetic structures. i. the commensurate case, Journal of Applied Crystallography 49~(5) (2016) 1750--1776.

\bibitem{xu2020high}
Y.~Xu, L.~Elcoro, Z.-D. Song, B.~J. Wieder, M.~Vergniory, N.~Regnault, Y.~Chen, C.~Felser, B.~A. Bernevig, High-throughput calculations of magnetic topological materials, Nature 586~(7831) (2020) 702--707.

\bibitem{xu2011inorganic}
Y.~Xu, M.~Yamazaki, P.~Villars, Inorganic materials database for exploring the nature of material, Japanese Journal of Applied Physics 50~(11S) (2011) 11RH02.

\bibitem{connolly2012bibliography}
T.~F. Connolly, Bibliography of magnetic materials and tabulation of magnetic transition temperatures, Springer Science \& Business Media, 2012.

\bibitem{buschow2003handbook}
K.~J. Buschow, Handbook of magnetic materials, Elsevier, 2003.

\bibitem{villars2004data}
P.~Villars, K.~Cenzual, J.~Daams, Y.~Chen, S.~Iwata, Data-driven atomic environment prediction for binaries using the mendeleev number: Part 1. composition ab, Journal of alloys and compounds 367~(1-2) (2004) 167--175.

\bibitem{swain2016chemdataextractor}
M.~C. Swain, J.~M. Cole, Chemdataextractor: a toolkit for automated extraction of chemical information from the scientific literature, Journal of chemical information and modeling 56~(10) (2016) 1894--1904.

\bibitem{court2018auto}
C.~J. Court, J.~M. Cole, Auto-generated materials database of curie and n{\'e}el temperatures via semi-supervised relationship extraction, Scientific data 5~(1) (2018) 1--12.

\bibitem{gilligan2023rule}
L.~P. Gilligan, M.~Cobelli, V.~Taufour, S.~Sanvito, A rule-free workflow for the automated generation of databases from scientific literature, arXiv preprint arXiv:2301.11689 (2023).

\bibitem{himanen2020dscribe}
L.~Himanen, M.~O. J{\"a}ger, E.~V. Morooka, F.~F. Canova, Y.~S. Ranawat, D.~Z. Gao, P.~Rinke, A.~S. Foster, Dscribe: Library of descriptors for machine learning in materials science, Computer Physics Communications 247 (2020) 106949.

\bibitem{vaswani2017attention}
A.~Vaswani, N.~Shazeer, N.~Parmar, J.~Uszkoreit, L.~Jones, A.~N. Gomez, {\L}.~Kaiser, I.~Polosukhin, Attention is all you need, Advances in neural information processing systems 30 (2017).

\bibitem{devlin2018bert}
J.~Devlin, M.-W. Chang, K.~Lee, K.~Toutanova, Bert: Pre-training of deep bidirectional transformers for language understanding, arXiv preprint arXiv:1810.04805 (2018).

\bibitem{brown2020language}
T.~Brown, B.~Mann, N.~Ryder, M.~Subbiah, J.~D. Kaplan, P.~Dhariwal, A.~Neelakantan, P.~Shyam, G.~Sastry, A.~Askell, et~al., Language models are few-shot learners, Advances in neural information processing systems 33 (2020) 1877--1901.

\bibitem{ouyang2022training}
L.~Ouyang, J.~Wu, X.~Jiang, D.~Almeida, C.~Wainwright, P.~Mishkin, C.~Zhang, S.~Agarwal, K.~Slama, A.~Ray, et~al., Training language models to follow instructions with human feedback, Advances in Neural Information Processing Systems 35 (2022) 27730--27744.

\bibitem{touvron2023llama2}
H.~Touvron, L.~Martin, K.~Stone, P.~Albert, A.~Almahairi, Y.~Babaei, N.~Bashlykov, S.~Batra, P.~Bhargava, S.~Bhosale, et~al., Llama 2: Open foundation and fine-tuned chat models, arXiv preprint arXiv:2307.09288 (2023).

\bibitem{deletang2023language}
G.~Del{\'e}tang, A.~Ruoss, P.-A. Duquenne, E.~Catt, T.~Genewein, C.~Mattern, J.~Grau-Moya, L.~K. Wenliang, M.~Aitchison, L.~Orseau, et~al., Language modeling is compression, arXiv preprint arXiv:2309.10668 (2023).

\bibitem{zhou2022large}
Y.~Zhou, A.~I. Muresanu, Z.~Han, K.~Paster, S.~Pitis, H.~Chan, J.~Ba, Large language models are human-level prompt engineers, arXiv preprint arXiv:2211.01910 (2022).

\bibitem{johnson2019billion}
J.~Johnson, M.~Douze, H.~J{\'e}gou, Billion-scale similarity search with {GPUs}, IEEE Transactions on Big Data 7~(3) (2019) 535--547.

\bibitem{wick2018calamari}
C.~Wick, C.~Reul, F.~Puppe, Calamari-a high-performance tensorflow-based deep learning package for optical character recognition, arXiv preprint arXiv:1807.02004 (2018).

\end{thebibliography}





\end{document}